\documentclass[12pt,reqno]{article}
\usepackage{amsfonts}
\usepackage{amsmath}
\usepackage{amsbsy}

\usepackage{amssymb,latexsym}
\numberwithin{equation}{section}

\begin{document}
 \allowdisplaybreaks[1]
\title{Dualisation of the Principal Sigma Model}
\author{Nejat T. Y$\i$lmaz\\
Department of Mathematics
and Computer Science,\\
\c{C}ankaya University,\\
\"{O}\u{g}retmenler Cad. No:14,\quad  06530,\\
 Balgat, Ankara, Turkey.\\
          \texttt{ntyilmaz@cankaya.edu.tr}}
\maketitle
\begin{abstract}
The first-order formulation of the principal sigma model with a
Lie group target space is performed. By using the dualisation of
the algebra and the field content of the theory the field
equations which are solely written in terms of the field strengths
are realized through an extended symmetry algebra parametrization.
The structure of this symmetry algebra is derived so that it
generates the realization of the field equations in a Bianchi
identity of the current derived from the extended parametrization.
\end{abstract}

\section{Introduction}
The dualisation or the first-order formulation method
\cite{julia2} of supergravity theories has been used in
\cite{nej1} and \cite{nej2} to construct the extended coset
realizations of the symmetric space sigma models
\cite{sm1,sm2,sm3}. Furthermore the dualisation of the symmetric
space sigma model with matter couplings is performed in
\cite{nej3}. In this work we consider the principal sigma model
\cite{sig1,sig2,sig3,witten,pcm} with a Lie Group target space in
its Lie algebra parametrization. By using the Lagrange multiplier
method which treats the field strengths of the scalar field
content of the theory as fundamental fields we will derive the first-order field
equations. Then by taking the exterior derivative of these
equations we will obtain the second-order field equations which
can be expressed only in terms of the field strengths. Afterwards
we will use these field strength containing field equations to
achieve the dualisation of the theory in which we double the field
content and extend the symmetry algebra. We will show that the
above-mentioned field equations can be realized through
introducing a parametrization generated by the doubled field
content and the extended symmetry algebra such that the
Cartan-form associating this extended parametrization leads us to
the second-order field equations of the field strengths by
satisfying the Cartan-Maurer equation which is a Bianchi identity.
From another point of view we will derive the structure of the
extended symmetry algebra so that the above picture is justified.
We will also denote that the first-order field equations can also
be obtained from the dualized Cartan-form via a twisted
self-duality condition which is satisfied by it. Section two is
reserved for the first-order formulation of the theory whereas
section three will present the dualisation.
\section{First-Order Formulation}
In this section starting from the Lagrangian of the principal
sigma model \cite{sig1,sig2,sig3,witten,pcm} by using the Lagrange
multiplier method we will derive the first-order field equations
of the theory. The Lagrangian of the principal sigma model whose
target space is a Lie group $G$ can be given as
\begin{equation}\label{de1}
 {\mathcal{L}}=\frac{1}{2}\, tr(\ast dg^{-1}\wedge
 dg).
\end{equation}
In the present work we will consider the exponential
parametrization of the group $G$ thus we will assume that the
group map $g$ in the above Lagrangian is the exponential map
\begin{equation}\label{de2}
g=e^{\varphi^{i}(x)T_{i}}.
\end{equation}
Specifically to define the above map we first introduce
\begin{equation}\label{de4}
h : M\longrightarrow G,
\end{equation}
which is a differentiable map from a $D$-dimensional spacetime $M$
into the group $G$. Then we consider a representation
\begin{equation}\label{de5}
f: G\longrightarrow Gl(N,\Bbb{R}),
\end{equation}
of $G$ in $Gl(N,\Bbb{R})$ for some $N$ which may be taken as a
differentiable homomorphism. Thus finally $g$ which is defined to
be the composition of \eqref{de4} and \eqref{de5} becomes a
matrix-valued function on $M$. In \eqref{de2} for
$i=1,\cdots,$dim$G$ $\{T_{i}\}$ are the generators of the Lie
algebra of $G$ and the fields $\{\varphi^{i}(x)\}$ are the scalars
on $M$. On the other hand the trace in \eqref{de1} is over the
matrix representatives of the matrix representation \eqref{de5}.
Now by using the matrix identities
\begin{equation}\label{de7}
g^{-1}dg=-dg^{-1}g\quad ,\quad dgg^{-1}=-gdg^{-1},
\end{equation}
we can express the Lagrangian \eqref{de1} in terms of the
invariant Noether current
\begin{equation}\label{de8}
\mathcal{G}^{\prime}=g^{-1}dg,
\end{equation}
as
\begin{equation}\label{de9}
{\mathcal{L}}=-\frac{1}{2}\, tr(\ast \mathcal{G}^{\prime}\wedge
\mathcal{G}^{\prime}).
\end{equation}
Again by making use of the identities \eqref{de7} one can show
that the Lie algebra valued Noether current
\begin{equation}\label{de9.5}
\mathcal{G}^{\prime}=F^{m}T_{m},
\end{equation}
satisfies the Bianchi identity
\begin{equation}\label{de10}
d\mathcal{G}^{\prime}+ \mathcal{G}^{\prime}\wedge
\mathcal{G}^{\prime}=0.
\end{equation}
At this stage we may adopt the formulation of \cite{nej2} where
the Noether current \eqref{de8} is explicitly calculated in terms
of the scalar fields $\{\varphi^{i}(x)\}$. It reads
\begin{equation}\label{de11}
\mathcal{G}^{\prime}=W^{m}_{\:\:\: n}d\varphi^{n}T_{m},
\end{equation}
where the dim$G\times$dim$G$ matrix $W$ can be given as
\begin{equation}\label{de12}
W=(I-e^{-M})M^{-1}.
\end{equation}
In the above expression the matrix $M$ is defined as
\begin{equation}\label{de13}
M_{\:\:\:m}^{n}=C_{lm}^{n}\varphi^{l},
\end{equation}
where we use the structure constants of the Lie algebra of $G$
defined via
\begin{equation}\label{de14}
[T_{m},T_{n}]=C_{mn}^{l}T_{l}.
\end{equation}
The coefficients of the algebra generators in \eqref{de11} become
elements of a standard kinetic term if \eqref{de11} is inserted in \eqref{de9}
thus we may simply call them the field strengths of the scalar fields
$\{\varphi^{i}(x)\}$ and they read
\begin{equation}\label{de15}
F^{m}=W^{m}_{\:\:\: n}d\varphi^{n}.
\end{equation}
The second-order field equations of the Lagrangian \eqref{de1} are
already obtained in terms of the scalar fields in \cite{sssm2}.
However in the following formulation for their relevant
advantageous role in the dualisation of the theory and for their
own right we will use the Lagrange multiplier method to derive the
locally-integrated first-order field equations of \eqref{de1}. For
this purpose we will focus on the field strengths \eqref{de15}.
Now if we insert \eqref{de11} in the Bianchi identity \eqref{de10}
we get
\begin{equation}\label{de16}
dF^{m}T_{m}+F^{m}\wedge F^{n}T_{m}T_{n}=0.
\end{equation}
By using the properties of the wedge product and also by doing index
redefinitions the second term in \eqref{de16} can be written as
\begin{equation}\label{de17}
\begin{aligned}
F^{m}\wedge F^{n}T_{m}T_{n}&=W^{m}_{\:\:\: k}W^{n}_{\:\:\: l}d\varphi^{k}\wedge d\varphi^{l}T_{m}T_{n}\\
\\
&=\frac{1}{2}F^{m}\wedge F^{n}[T_{m},T_{n}]\\
\\
&=\frac{1}{2}C_{mn}^{l}F^{m}\wedge F^{n}T_{l}.
\end{aligned}
\end{equation}
After inserting this result back in \eqref{de16} if we equate the
coefficients of the generators which are linearly independent to zero we
can write the Bianchi identity in terms of the field strengths as
\begin{equation}\label{de18}
dF^{l}=-\frac{1}{2}C_{mn}^{l}F^{m}\wedge F^{n}.
\end{equation}
To derive the first-order field equations of the Lagrangian
\eqref{de1} instead of the scalar fields $\{\varphi^{i}(x)\}$ we
will treat their field strengths \eqref{de15} which satisfy the
Bianchi identities \eqref{de18} as independent fields. For this reason we
will construct a Bianchi Lagrangian starting from the identities in \eqref{de18} which can be
considered as constraint relations on the field strengths.
Now for each $F^{l}$ after introducing a Lagrange
multiplier $A_{l}$ which is a $(D-2)$-form we can write the
Bianchi Lagrangian as
\begin{equation}\label{de19}
\mathcal{L}_{Bianchi}=(dF^{l}+\frac{1}{2}C_{mn}^{l}F^{m}\wedge
F^{n})\wedge A_{l}.
\end{equation}
As we have mentioned above the Lagrangian \eqref{de1} can also be written in terms of the field
strengths as
\begin{equation}\label{de20}
\mathcal{L}=-\frac{1}{2}\ast F^{m}\wedge F^{n}T_{mn},
\end{equation}
where we introduce the trace convention of the matrix
representatives of the generators as
\begin{equation}\label{de21}
T_{mn}=tr(T_{m}T_{n}),
\end{equation}
which is obviously symmetric. The total Lagrangian in which the
field strengths replace the scalars and in which they can be treated as independent fields can now be
written as
\begin{equation}\label{de22}
\mathcal{L}_{tot}=-\frac{1}{2}\ast F^{m}\wedge
F^{n}T_{mn}+(dF^{l}+\frac{1}{2}C_{mn}^{l}F^{m}\wedge F^{n})\wedge
A_{l}.
\end{equation}
If we vary this Lagrangian with respect to the Lagrange
multipliers $A_{l}$ we obtain the Bianchi identities \eqref{de18}.
However varying the above Lagrangian with respect to $F^{l}$ will
give us the desired first-order field equations. If we do so the
corresponding field equations of $F^{l}$ from \eqref{de22} can be
derived as
\begin{equation}\label{de23}
(-1)^{D}T_{ml}\ast F^{m}=-dA_{l}-C_{ln}^{k}F^{n}\wedge A_{k}.
\end{equation}
Clearly these equations are first-order in terms of the
scalar fields $\{\varphi^{i}(x)\}$ and the $(D-2)$-forms $A_{l}$
are completely arbitrary so that one can run them on the entire
set of $(D-2)$-forms to obtain the general solutions of the theory. In
the next section we will also need the second-order field
equations solely in terms of the field strengths $F^{l}$ which have a
key role in the dualisation of the theory. For this reason we will also derive them here
by taking the exterior derivative of \eqref{de23}. If we apply the
exterior derivative on both sides of \eqref{de23} we find
\begin{equation}\label{de24}
\begin{aligned}
(-1)^{D}T_{ml}d(\ast F^{m})&=-(-1)^{D}C_{ln}^{k}F^{n}\wedge\ast
F^{m}T_{mk}+\frac{1}{2}C_{ln}^{k}C_{uv}^{n}F^{u}\wedge
F^{v}\wedge A_{k}\\
&\quad-C_{ln}^{k}C_{kv}^{t}F^{n}\wedge F^{v}\wedge A_{t},
\end{aligned}
\end{equation}
where we have used the Bianchi identities \eqref{de18} and the
first-order field equations \eqref{de23} again. Now by using the Jacobi
identities
\begin{equation}\label{de25}
C_{ln}^{k}C_{uv}^{n}=-C_{vl}^{n}C_{un}^{k}-C_{lu}^{n}C_{vn}^{k},
\end{equation}
after some algebra one can show that
\begin{equation}\label{de26}
\frac{1}{2}C_{ln}^{k}C_{uv}^{n}F^{u}\wedge F^{v}\wedge
A_{k}=C_{ln}^{k}C_{kv}^{t}F^{n}\wedge F^{v}\wedge A_{t}.
\end{equation}
Thus the last two terms on the right hand side of \eqref{de24}
cancel out and in terms of the field strengths $F^{l}$ we have the
final form of the second-order field equations
\begin{equation}\label{de27}
d(T_{ml}\ast F^{m})=-C_{ln}^{k}T_{mk}F^{n}\wedge\ast F^{m}.
\end{equation}
\section{Dualisation}
In this section we will follow the dualisation method of
\cite{julia2} to construct an enlarged exponential parametrization
of the theory which will give us the first-order field equations
\eqref{de23} in a geometrical formalism. We start by introducing
the dual fields $\{\widetilde{\varphi}^{l}\}$ which are
$(D-2)$-forms. We will also introduce the dual generators
$\{\widetilde{T}_{l}\}$ which will couple to the dual fields in
the extended exponential map. We will assume that the generators
\begin{equation}\label{de28}
\{T_{l},\widetilde{T}_{l}\},
\end{equation}
form up an algebra. Of course from \eqref{de14} we know the
commutation relations of the generators $\{T_{l}\}$ which
constitute a sub-Lie algebra in the extended algebra generated by
\eqref{de28}. We will assume that the algebra of the generators
\eqref{de28} has a $\mathbb{Z}_{2}$ grading. The generators are
chosen to be odd if the corresponding coupling field is an odd
degree differential form and otherwise even. Thus $\{T_{l}\}$ are
even generators however $\{\widetilde{T}_{l}\}$ are even or odd
depending on the dimension $D$ of $M$. In the construction of the
extended exponential map which will be based on the generators
\eqref{de28} we will make use of the differential graded algebra
structure of the differential forms and the field generators.
Within the differential graded algebra structure the odd (even)
generators behave like odd (even) degree differential forms when
they commute with the exterior product. The algebra products of
the odd generators obey the anti-commutation relations on the
other hand the algebra products of the even ones and the mixed
ones obey the commutation relations. Now let us define the map
\begin{equation}\label{de29}
g^{\prime}=e^{\varphi^{i}T_{i}}e^{\widetilde{\varphi}^{j}\widetilde{T}_{j}},
\end{equation}
and consider the Cartan-Maurer form
\begin{equation}\label{de30}
\mathcal{G}^{\prime\prime}=(g^{\prime})^{-1}dg^{\prime}.
\end{equation}
Our aim will be to derive the algebra structure of the generators
\eqref{de28} so that when the Cartan-Maurer form \eqref{de30} is
inserted in the Cartan-Maurer equation
\begin{equation}\label{de31}
d\mathcal{G}^{\prime\prime}+ \mathcal{G}^{\prime\prime}\wedge
\mathcal{G}^{\prime\prime}=0,
\end{equation}
which is a Bianchi identity for $\mathcal{G}^{\prime\prime}$ it yields the second-order field equations \eqref{de27}. Now
after substituting
\eqref{de29} in \eqref{de30} we get
\begin{equation}\label{de32}
\mathcal{G}^{\prime\prime}=e^{-\widetilde{\varphi}^{j}\widetilde{T}_{j}}\mathcal{G}^{\prime}e^{\widetilde{\varphi}^{j}\widetilde{T}_{j}}+
e^{-\widetilde{\varphi}^{j}\widetilde{T}_{j}}de^{\widetilde{\varphi}^{j}\widetilde{T}_{j}}.
\end{equation}
By using the matrix identities
\begin{equation}\label{de33}
\begin{aligned}
e^{-C} de^{C}&=dC-\frac{1}{2!}[C,dC]+\frac{1}{3!}[C,[C,
dC]]-\cdots,\\
\\
e^{-X}Ye^{X}&=Y-[X,Y]+\frac{1}{2!}[X,[X,Y]]-\cdots,
\end{aligned}
\end{equation}
one can further simplify \eqref{de32} however we will postpone
this derivation at this moment and before dealing with it we will mention about
some crucial remarks on the algebra structure and on $\mathcal{G}^{\prime\prime}$. If we follow the above-mentioned
track of deriving the algebra structure at first glance we observe that to obtain the
correct form of field equations from \eqref{de31} we must have
\begin{equation}\label{de34}
[\widetilde{T}_{j},\widetilde{T}_{k}\}=0.
\end{equation}
This is a general feature of the dualisation of the field
theories. Another fact about the doubled-formalism of the field theories
is that we should require that $\mathcal{G}^{\prime\prime}$ will obey a twisted
self-duality condition
\begin{equation}\label{de35}
\ast\mathcal{G}^{\prime\prime}=\mathcal{S}\mathcal{G}^{\prime\prime}.
\end{equation}
This condition results in the first-order field equations
\eqref{de23} of the theory. Here $\mathcal{S}$ is a
pseudo-involution of the generators in \eqref{de28}. In general
$\mathcal{S}$ maps the original generators onto the dual ones and
the dual generators are mapped onto the original ones with a sign factor. When applied twice
on the original generators it has the same eigenvalues $\pm1$ with the action of the operator
$\ast\cdot\ast$ on the corresponding dual field strengths of the
dual potentials \cite{julia2}. Thus
$\mathcal{S}\widetilde{T}_{i}=(-1)^{(p(D-p)+s)}T_{i}$ where $p$ is
the degree of the dual field strength and $s$ is the signature of
the spacetime. The degree of our dual field strengths
corresponding to the dual generators is $(D-1)$ and if we also
assume that the signature of $M$ is $s=1$ then we have
\begin{equation}\label{de36}
\mathcal{S}T_{i}=\widetilde{T}_{i}\quad
,\quad\mathcal{S}\widetilde{T}_{i}=(-1)^{D}T_{i}.
\end{equation}
Now by using \eqref{de33} and \eqref{de34} from \eqref{de32} we
have
\begin{equation}\label{de37}
\mathcal{G}^{\prime\prime}=\mathcal{G}^{\prime}-[\widetilde{\varphi}^{j}\widetilde{T}_{j},\mathcal{G}^{\prime}]
+d\widetilde{\varphi}^{j}\widetilde{T}_{j}.
\end{equation}
We can write $\mathcal{G}^{\prime\prime}$ in the form
\begin{equation}\label{de38}
\mathcal{G}^{\prime\prime}=F^{m}T_{m}+\widetilde{F}^{j}\widetilde{T}_{j}.
\end{equation}
If we use this expression in \eqref{de35} with the help of
\eqref{de36} we see that
\begin{equation}\label{de39}
\widetilde{F}^{i}=(-1)^{D}\ast F^{i}.
\end{equation}
Thus the on-shell form of $\mathcal{G}^{\prime\prime}$ must be
\begin{equation}\label{de40}
\mathcal{G}^{\prime\prime}=F^{m}T_{m}+(-1)^{D}\ast
F^{i}\widetilde{T}_{i}.
\end{equation}
This form of the doubled current which is solely written in terms
of the original field strengths $F^{l}$ when inserted in
\eqref{de31} should lead us to the second-order field equations
\eqref{de27}. In this way we can determine the structure constants
of the algebra generated by \eqref{de28}. If we perform the above
mentioned substitution we get
\begin{equation}\label{de41}
\begin{aligned}
0&=dF^{m}T_{m}+F^{m}T_{m}\wedge F^{n}T_{n}+(-1)^{D}d(\ast F^{i})\widetilde{T}_{i}\\
\\
&\quad +(-1)^{D}F^{m}T_{m}\wedge\ast
F^{i}\widetilde{T}_{i}+(-1)^{D}\ast F^{i}\widetilde{T}_{i}\wedge
F^{m}T_{m}\\
\\
&\quad+\ast F^{i}\widetilde{T}_{i}\wedge\ast
F^{j}\widetilde{T}_{j}.
\end{aligned}
\end{equation}
The first two terms are automatically zero from \eqref{de10}. If
we assume that $D>2$ the last term also vanishes. Now if we define
\begin{equation}\label{de42}
[T_{m},\widetilde{T}_{i}]=D_{mi}^{l}\widetilde{T}_{l},
\end{equation}
and use the fact that the generator $\{\widetilde{T}_{l}\}$ are
linearly independent from \eqref{de41} we read
\begin{equation}\label{de43}
d(\ast F^{l})=-D_{mi}^{l}F^{m}\wedge\ast F^{i}.
\end{equation}
Multiplying both sides by $T_{ml}$ and summing on the index $m$ gives
\begin{equation}\label{de44}
d(T_{ml}\ast F^{m})=-T_{kl}D_{nm}^{k}F^{n}\wedge\ast F^{m}.
\end{equation}
Now comparing this result with the second-order field equations
\eqref{de27} reveals the desired structure constants. From this comparison we find that
\begin{equation}\label{de45}
T_{kl}D_{nm}^{k}=C_{ln}^{k}T_{mk}.
\end{equation}
Now to extract the structure constants from \eqref{de45} if we define the matrices $T,D_{n},C_{n}$ as
\begin{equation}\label{de46}
(T)^{l}_{\:\:\:k}=T_{lk}\quad ,\quad
(D_{n})^{l}_{\:\:\:k}=D_{nk}^{l}\quad ,\quad
(C_{n})^{l}_{\:\:\:k}=C_{nk}^{l},
\end{equation}
then we have
\begin{equation}\label{de47}
D_{n}=-T^{-1}C_{n}^{T}T\quad ,\quad
D_{nk}^{l}=-(T^{-1}C_{n}^{T}T)^{l}_{\:\:\:k},
\end{equation}
which completely determines the remaining structure constants of
the doubled symmetry algebra. Finally to recover the first-order
field equations if we insert \eqref{de37} in \eqref{de35} after
some algebra by also using \eqref{de42} we get
\begin{equation}\label{de48}
(-1)^{D}\ast F^{l}=D_{mj}^{l}F^{m}\wedge
\widetilde{\varphi}^{j}+d\widetilde{\varphi}^{l}.
\end{equation}
By multiplying both sides with $T_{ml}$ and summing on $m$ then also
by using \eqref{de45} we find that
\begin{equation}\label{de49}
(-1)^{D}T_{ml}\ast F^{m}=C_{lm}^{n}T_{jn}F^{m}\wedge
\widetilde{\varphi}^{j}+T_{nl}d\widetilde{\varphi}^{n}.
\end{equation}
We realize that under the field redefinitions
\begin{equation}\label{de50}
A_{n}=-T_{jn}\widetilde{\varphi}^{j},
\end{equation}
\eqref{de49} and \eqref{de23} are the same equations. This is a
legitimate result since the dual fields or the Lagrange
multipliers are auxiliary fields which arise from the local
integration or the first-order formulation of the theory.
\section{Conclusion}
After deriving the Bianchi identities for the field strengths of
the scalar fields which constitute the principal sigma model
Lagrangian for a Lie group target space by parametrizing the exponential map we
have considered these identities as constraints on the field
strengths and by introducing Lagrange multipliers we have
constructed the Bianchi Lagrangian. Treating the field strengths
as independent fields we have derived the first-order field
equations of the theory after adding the Bianchi Lagrangian to the
original one. We have also obtained the second-order field
equations only in terms of the field strengths by differentiating
these first-order equations. These first and second-order field equations
are then used as building blocks of the dualisation of the theory
in section three where we have discovered the structure of the
extended symmetry algebra which realizes these equations in a
geometrical formalism. To do this we have introduced dual fields
and dual generators and doubled the content of the theory. After
introducing an extended parametrization map we have derived the
structure constants of the dual generators so that the Cartan-form
of the new map generates the field equations of the theory in the
Cartan-Maurer equation. In this context the first-order field
equations are also obtained from a twisted self-duality equation
of the Cartan-form.

Our formulation extends the results of \cite{julia2} to the most
general class of principal sigma models. We have studied these
models in their generic form. Therefore we have obtained results
containing arbitrary structure constants and trace conventions.
The dualisation of the principal sigma model completes a
discussion given in \cite{julia2} which points to the
enhanced symmetry study of the principal sigma model by dualizing it. In this direction
depending on the extended algebra structure, the first and, the second-order field equations
of the present paper one may investigate the extended symmetries  of the general principal sigma model. The symmetries of both
the first and the second-order field equations derived here can also be
studied and they can be compared with the extended symmetry
algebra.

In section three, although
we have built our analysis on a dualized parametrization map we have not questioned its
geometrical meaning. We have simply considered this map as a
differential form valued matrix which is a legitimate consequence of the matrix representation pre-assumed for the
extended symmetry algebra.
One may furthermore work on the
geometrical and the group theoretical aspects of this extended
parametrization we have introduced to dualize the theory. Such an attempt would reveal
a more generalized geometrical construction of the theory at the dualisation level. As a
final remark we should state that apart from the dualisation of the
general principal sigma model with a Lie group target space this
work presents the first-order formulation of the theory which is
an important result in its own respect. The first-order field equations of the theory which are written in their generic form
can provide essential tools in finding solutions even when the theory is coupled to various other fields.

\end{document}